\documentclass[review,12pt,authoryear]{elsarticle}
\usepackage[utf8]{inputenc}
\usepackage[T1]{fontenc}
\usepackage{lineno}

\usepackage{url}
\usepackage{amsmath}
\usepackage{graphicx}
\usepackage{geometry}
\usepackage{booktabs}
\usepackage{algorithm}
\usepackage{algpseudocode}
\usepackage{setspace}
\journal{Elsevier}

\begin{document}

\begin{frontmatter}

\title{Joint Analysis of Shannon and Tsallis Entropy and GRACE-FO driven Equivalent Water Height Anomalies for Pre- and Post-Rupture Monitoring: An Example of the 2023 $M_w = 7.8$ Kahramanmara\c{s} Earthquake, T\"{u}rkiye}

\author[1]{Muhammed Hossein Mousavi}
\author[2]{Hamzeh Mohammadigheymasi\corref{cor1}}
\ead{hamzeh.prof@gmail.com}
\author[3]{Amirreza Moradi}

\cortext[cor1]{Corresponding author}

\address[1]{Department of Physics Education, Farhangian University, P.O. Box 14665-889, Tehran, Iran}
\address[2]{Atmosphere and Ocean Research Institute, The University of Tokyo, Kashiwa, Chiba, Japan}
\address[3]{Department of Geoscience Engineering, Arak University of Technology, Arak, Iran}

\begin{abstract}
In order to understand the variations in fault systems throughout the seismogenic cycle, mechanical states and the complexities of seismic interactions must be considered. In this study, we present a data integration framework combining a 25-year seismic catalog with Equivalent Water Height (EWH) datasets from the GRACE-FO mission and two information-theoretic complexity measures (Shannon and Tsallis entropy) to examine spatiotemporal changes in the East Anatolian Fault System associated with the 2023 Kahramanmara\c{s} earthquake doublet. The pre-rupture period exhibits a systematic increase in the entropy measures alongside a gradual decrease in EWH, suggesting a transition towards fault network criticality driven by segment fragmentation, long-range correlations, poroelastic contraction, fluid migration, and progressive stress accumulation. During the co-seismic phase, we observe an abrupt increase in entropy with a corresponding negative shift in EWH. In the post-seismic period, the persistence of elevated entropy and EWH anomalies indicates that the fault system remains in a non-equilibrium state dominated by aftershock clustering, fault zone damage, permeability changes, and viscoelastic relaxation. Additionally, structured computational workflows detailing these joint methodologies are provided via the Seismic Entropy Analysis (Algorithm 1) and the Relationship Between Tsallis $q$ and Gutenberg--Richter $b$-value (Algorithm 2) pseudo-codes, facilitating the direct reproduction and regeneration of all results.
\end{abstract}


\end{frontmatter}

\section{Introduction}\label{sec:intro}

Earthquakes represent abrupt and irreversible changes between different states of mechanical equilibrium, driven by the long-term accumulation and sudden release of elastic strain energies in the Earth's crust. Investigating these rupture characteristics and the underlying physical processes typically relies on indirect measurements obtained by analyzing geophysical datasets, most notably seismic catalogs. Despite significant progress in the field of observational seismology and development of high-resolution seismic catalogs \citep{mohammadigheymasi2023ipiml, carvalho2025application, khurshid2025comprehensive, mohammadigheymasi2023application}, the characterization of non-linear changes in complex fault systems is still a central problem in earthquake science. Because of the inability to measure subsurface stress directly, seismologists must rely on statistical patterns and surface observations to infer changes in the physical state of fault systems \citep{Scholz_2019, Kanamori_2004, Zoback_2010}. The classical approach of using the Gutenberg-Richter $b$-value and temporal changes in the rate of seismicity ($Z$-value) is widely used to estimate changes in stress concentrations, fault zone locking, and changes in seismic hazard \citep{Mogi_1967, Wyss1988, scholz2015, ozturk2018}. However, such methods may only give a partial description of the underlying dynamics. An entropy-theoretic approach offers a complementary framework for quantifying the disorder, complexity, and structural reorganization of seismic systems. Shannon Entropy \citep{Shannon_1948} has been widely used to estimate the unpredictability and spatiotemporal variability of earthquake systems \citep{De_Santis_2011, Vogel_2020}. Changes in entropy are considered an indication of the progression of the system to criticality, where long-range interactions are enhanced, and irreversible changes such as large earthquakes are likely \citep{Hopfield_1994, Herz_1995, Rundle2002}. Besides classical Boltzmann--Gibbs statistics, non-extensive statistics based on Tsallis entropy has been successfully employed to capture long-range correlations, memory effects, and multi-scale interactions, which are ubiquitous in seismicity in heterogeneous crustal environments \citep{Sotolongo_Costa_2004, Telesca_2010, Telesca_2011, Varotsos_2011, Posadas_2023}. The entropic index $q$, which controls the level of non-extensivity, has been related to the $b$-value of the Gutenberg-Richter law, providing a physical connection between statistical seismology and non-extensive thermodynamics. This connection allows a deeper analysis of the evolution of correlations in fault systems. At the same time, satellite gravimetry missions such as GRACE and GRACE-FO have made it possible to detect changes in the large-scale mass redistribution due to tectonic and hydrological processes. These missions have demonstrated their sensitivity to co-seismic effects, such as changes in gravity due to earthquakes, as well as viscoelastic relaxation and long-term deformation of the Earth's surface \citep{Han_2006, Panet_2007, Wang_2012}. In addition to gravimetric satellites, remote sensing mapping utilizing multispectral and radar sensors provides valuable lithological and structural constraints on active tectonic zones \citep{Wubneh2025}. Theoretical and observational results indicate that strong earthquakes ($M_w \ge 7.5$) are able to produce detectable changes in the Earth's gravity \citep{Sun_2004, Tesmer_2011, Tanaka_2015, Zou_2015, Pan_2016}. The GRACE-FO-derived Equivalent Water Height (EWH) anomalies are thus seen to offer another geodetic constraint on poroelastic response, fluid redistribution, and mass changes within the Earth's crust. The $M_w$ 7.8 and $M_w$ 7.7 Kahramanmara\c{s} earthquake doublet of 6 February 2023 occurred along the East Anatolian Fault Zone (EAFZ), a major left-lateral strike-slip fault accommodating the westward extrusion of the Anatolian microplate \citep{Barazangi_2006, G_vercin_2022, Rodr_guez_P_rez_2024}. These earthquakes have ruptured a longstanding seismic gap with a slip deficit of $\sim$5.2~m, providing a unique opportunity to study the preparatory, co-seismic, and post-seismic processes of a complex multi-fault system \citep{Ambraseys_1998, Ambraseys_2009}. This raises a critical, open scientific question: *Can the integration of information-theoretic complexity metrics and satellite-derived mass changes identify precursory criticality, map co-seismic strain release, and characterize post-seismic relaxation states in a complex, multi-fault strike-slip system?*

To test this hypothesis, we utilize the 2023 Kahramanmara\c{s} earthquake doublet as a natural laboratory. We address this question by establishing a joint analytical framework that tracks Shannon and Tsallis entropy variations derived from a 25-year regional seismic catalog alongside monthly terrestrial mass changes (EWH anomalies) from the GRACE-FO mission. This integrated approach allows us to determine whether statistical disorder in regional seismicity correlates with poroelastic fluid and mass redistribution during different phases of the seismogenic cycle, thereby contributing to more robust and physics-based seismic hazard assessment strategies.

The remainder of this paper is structured as follows. Section 2 outlines the theoretical formulation of the information-theoretic complexity measures, including Shannon and Tsallis entropy. Section 3 describes the seismic catalog and satellite gravimetry datasets used in this study. Section 4 presents the spatiotemporal analysis results and discusses their physical implications for the Kahramanmara\c{s} earthquake doublet. Finally, Section 5 summarizes the primary conclusions of this work.

\section{Methodology}\label{sec:method}

This section establishes the mathematical and statistical-mechanical foundations for modeling active tectonic fault systems as out-of-equilibrium dynamical networks. To diagnose the evolution of structural complexity and detect precursory signs of criticality, we implement a dual information-theoretic framework. First, we outline the Shannon entropy, which acts as a macroscopic measure of statistical disorder and phase space occupancy, revealing the information gain or loss in the spatial and temporal distributions of seismicity. Second, we formulate the non-extensive Tsallis entropy, which generalizes classical Boltzmann--Gibbs statistical mechanics to systems dominated by long-range correlations, fractal geometries, and memory effects. Together, these measures enable a rigorous quantification of the thermodynamic and statistical state transitions within the East Anatolian Fault Zone throughout the seismogenic cycle.

\subsection{The Shannon Entropy}\label{subsec:shannon}

In 1948, Claude Shannon introduced a pivotal concept of entropy within the realm of information theory \citep{Shannon_1948}. This concept pertains to the quantification of information content or the uncertainty present in a probabilistic framework. In a scenario characterized by a discrete array of potential states, where each state i is associated with a probability $p_i$, the Shannon entropy $H$ is articulated as follows:
\begin{equation}
    H = -\sum p_i \ln p_i
    \label{eq:shannon}
\end{equation}
with the additional condition that $p_i > 0$ for all i and $\sum p_i = 1$

This framework has been effectively applied in statistical physics for various analyses. However, a significant challenge in its application lies in the accurate estimation of the probabilities $p_i$. While direct computation is feasible for smaller systems (\citet{Negrete_2018}), the majority of real-world systems necessitate extensive sampling. Typically, this involves conducting N independent observations or samplings, counting the occurrences of each state i ($n_i$), and estimating the probability as follows:
\begin{equation}
    p_i \approx \frac{n_i}{N}
    \label{eq:shannon_est}
\end{equation}

A prevalent technique involves organizing the sampled data into a vector and creating a frequency histogram. By normalizing the count in each histogram bin against the total number of samples N, one can derive an empirical estimate of the probability distribution. Subsequently, this estimated distribution can be substituted into Eq. (\ref{eq:shannon}) to calculate the Shannon entropy. There exist alternative estimation techniques that ultimately converge on the fundamental principle of estimating the normalized likelihood of observable outcomes \citep{Shannon_1948, Posadas_2021}. The methodology employed in this paper utilizes the widely recognized histogram-based approach. The concept of entropy is extended from discrete variables to continuous variables. Consider a system defined by a continuous probability distribution with a density function $f(x)$. In this case, the definition of Shannon entropy is articulated through the concept of differential entropy:
\begin{equation}
    H = -\int f(x) \ln f(x) \, dx
    \label{eq:diff_entropy}
\end{equation}
where the integral is computed over all x for which $f(x)$ is defined, subject to the conditions $f(x) \ge 0$ and $\int f(x) \, dx = 1$. A significant finding in information theory that demonstrates the practical importance of entropy is the asymptotic equipartition property (AEP). This principle asserts that for a collection of n independent and identically distributed samples drawn from the distribution p, there exists a typical set T $\subseteq$ $S^n$ such that: 1. The likelihood that a sampled sequence is part of T approaches 1 as $n \to \infty$. 2. The number of elements in T is bounded above by $e^{n(H+\varepsilon)}$, where $\varepsilon$ is a small positive number that approaches 0 for large n. This implies that for large sample sizes, the effective number of probable states is exponentially governed by the Shannon entropy, solidifying its role as a measure of information and uncertainty.

\subsection{Tsallis Entropy}\label{subsec:tsallis}

In the realms of information theory and statistical mechanics, the Boltzmann-Gibbs-Shannon entropy has been extended through the introduction of Tsallis entropy, as proposed by Constantino Tsallis \citep{Tsallis_1988}. This extension aims to address the complexities of long-range interacting systems that exhibit memory effects or fractal characteristics, where traditional extensive thermodynamics fails to apply \citep{Telesca_2010, Telesca_2011, Varotsos_2011}. For a probability distribution $p = (p_1, p_2, p_3, \dots, p_\Omega)$ defined over a finite set of $\Omega$ possible states, the Tsallis entropy of order q (commonly represented by the entropic index $q$, which corresponds to $\alpha$ in the original text) is articulated as:
\begin{equation}
    S_q = \frac{1}{q-1} \left( 1 - \sum_{i=1}^{W} p_i^q \right)
    \label{eq:tsallis}
\end{equation}
where the sum encompasses all $\Omega$ potential states of the ensemble (for instance, amplitudes of seismic activities). A distinctive feature of this entropy is its non-additive nature for statistically independent systems when $q \neq 1$. In the limit as $q \to 1, S_1 = - \sum p_i \ln p_i$, hence justifying it as a proper generalization. In the field of seismology, the entropic index $q$ is not a random parameter but is fundamentally associated with the established Gutenberg-Richter (GR) law, which delineates the distribution of earthquakes in terms of frequency and magnitude. The $b$-value, representing the slope of the GR law, exhibits a relationship with the entropic index $q$ as follows:
\begin{equation}
    b = 2 \left( \frac{2-q}{q-1} \right)
    \label{eq:b_q_relation}
\end{equation}
This relationship typically restricts q to the range $1 < q \le 2$in the context of seismic phenomena. A value of $q > 1$ signifies a non-extensive system, wherein the seismicity is marked by strong correlations and long-range interactions among events. The deviation from the extensive Boltzmann-Gibbs condition (q=1) is noteworthy, as it suggests that the statistical mechanics governing earthquake activities cannot be fully explained through classical additive frameworks. Therefore, the Tsallis system offers a more appropriate statistical characterization for the intricate and dynamically correlated nature of the seismogenesis process. All entropy calculations (Shannon and Tsallis) were performed on a uniform spatial grid with a step size of $0.25^\circ$, which was used consistently for constructing all probability distributions.

\subsection{Computational Workflow}\label{subsec:workflow}

To systematically compute the spatial and temporal variations in Shannon and Tsallis entropy across the study region, we implement the unified processing pipeline detailed in Algorithm 1. Similar advancements in geostatistical modeling and spatial prediction emphasize the integration of statistical estimators to improve the mapping of environmental and geophysical variables \citep{Korniyenko2026}.

\begin{algorithm}[htbp]
\singlespacing
\caption{Seismic Entropy Analysis (Spatial \& Temporal Shannon / Tsallis)}
\label{alg:entropy_analysis}
\begin{algorithmic}[1]\footnotesize
\Require{catalog CSV \{time, latitude, longitude, magnitude\}, $\Delta m$, $\Delta \text{lon}$, $\Delta \text{lat}$, $W$ (window), $S$ (step), $N_{\text{min}}$}
\Statex \textbf{Phase 1 --- Global Parameters}
\State $\text{bins} \gets \text{ARANGE}(\min(M), \max(M)+\Delta m, \Delta m)$, $\text{counts} \gets \text{HISTOGRAM}(M, \text{bins})$
\State $M_c \gets \text{centre of bin with } \text{ARGMAX}(\text{counts})$
\State $M_{\text{filtered}} \gets M[M \ge M_c]$, $M_{\text{min}} \gets M_c - \Delta m/2$
\State $b \gets \log_{10}(e) / (\text{MEAN}(M_{\text{filtered}}) - M_{\text{min}})$, $q \gets (b + 4) / (b + 2)$
\Statex \textbf{Phase 2 --- Entropy Functions}
\State \textbf{function} \text{Shannon}(p): \Return $-\sum_{i} p_i \log_2(p_i)$
\State \textbf{function} \text{Tsallis}(p, q): \Return $|q - 1| < \epsilon$ ? \text{Shannon}(p) : $(1 - \sum_{i} p_i^q)/(q - 1)$
\Statex \textbf{Phase 3 --- Spatial Entropy Map}
\For{\textbf{each} cell $(\text{lon}_c, \text{lat}_c)$ \textbf{in} $\text{GRID}(\text{lon}_{\text{edges}}, \text{lat}_{\text{edges}})$}
    \State $\text{events} \gets \text{catalog}[\text{lon} \in \text{lon}_{\text{bin}} \land \text{lat} \in \text{lat}_{\text{bin}}]$
    \If{$|\text{events}| \ge N_{\text{min}}$}
        \State $p \gets \text{HISTOGRAM}(\text{events}[\text{magnitude}]) / |\text{events}|$
        \State $S_H(\text{lon}_c, \text{lat}_c) \gets \text{Shannon}(p)$, $S_T(\text{lon}_c, \text{lat}_c) \gets \text{Tsallis}(p, q)$
    \EndIf
\EndFor
\Statex \textbf{Phase 4 --- Temporal Entropy Series}
\State $t_0 \gets \min(\text{time})$, $t_{\text{end}} \gets \max(\text{time})$
\While{$t_0 + W \le t_{\text{end}}$}
    \State $\text{window} \gets \text{catalog}[t_0 \le \text{time} < t_0 + W]$
    \If{$|\text{window}| \ge N_{\text{min}}$}
        \State $p \gets \text{HISTOGRAM}(\text{window}[\text{magnitude}]) / |\text{window}|$, $t_{\text{mid}} \gets t_0 + W/2$
        \State Append $(t_{\text{mid}}, \text{Shannon}(p))$ to $H_{S,\text{series}}$, $(t_{\text{mid}}, \text{Tsallis}(p, q))$ to $H_{T,\text{series}}$
    \EndIf
    \State $t_0 \gets t_0 + S$
\EndWhile
\Statex \textbf{Output}
\State Write files: $\text{spatial\_entropy.csv} \gets \{ S_H, S_T \}$, $\text{temporal.csv} \gets \{ H_{S,\text{series}}, H_{T,\text{series}} \}$
\State Print $M_c, b, q$
\end{algorithmic}
\end{algorithm}

By establishing these mathematical frameworks and computational workflows, we are equipped to quantify both the additive, short-range informational uncertainty (Shannon entropy) and the non-additive, long-range correlations (Tsallis entropy) of seismic events. To apply these theoretical formulations to the 2023 Kahramanmara\c{s} earthquake doublet, we require high-resolution seismicity catalogs and geodetic measurements of crustal deformation. The specific datasets, preprocessing protocols, and spatial resolutions utilized in this study are detailed in Section 3.

\section{Dataset}\label{sec:dataset}

\subsection{Seismic Data}\label{subsec:seismic}

\begin{figure}[ht!]
    \centering
    \includegraphics[width=0.99\textwidth]{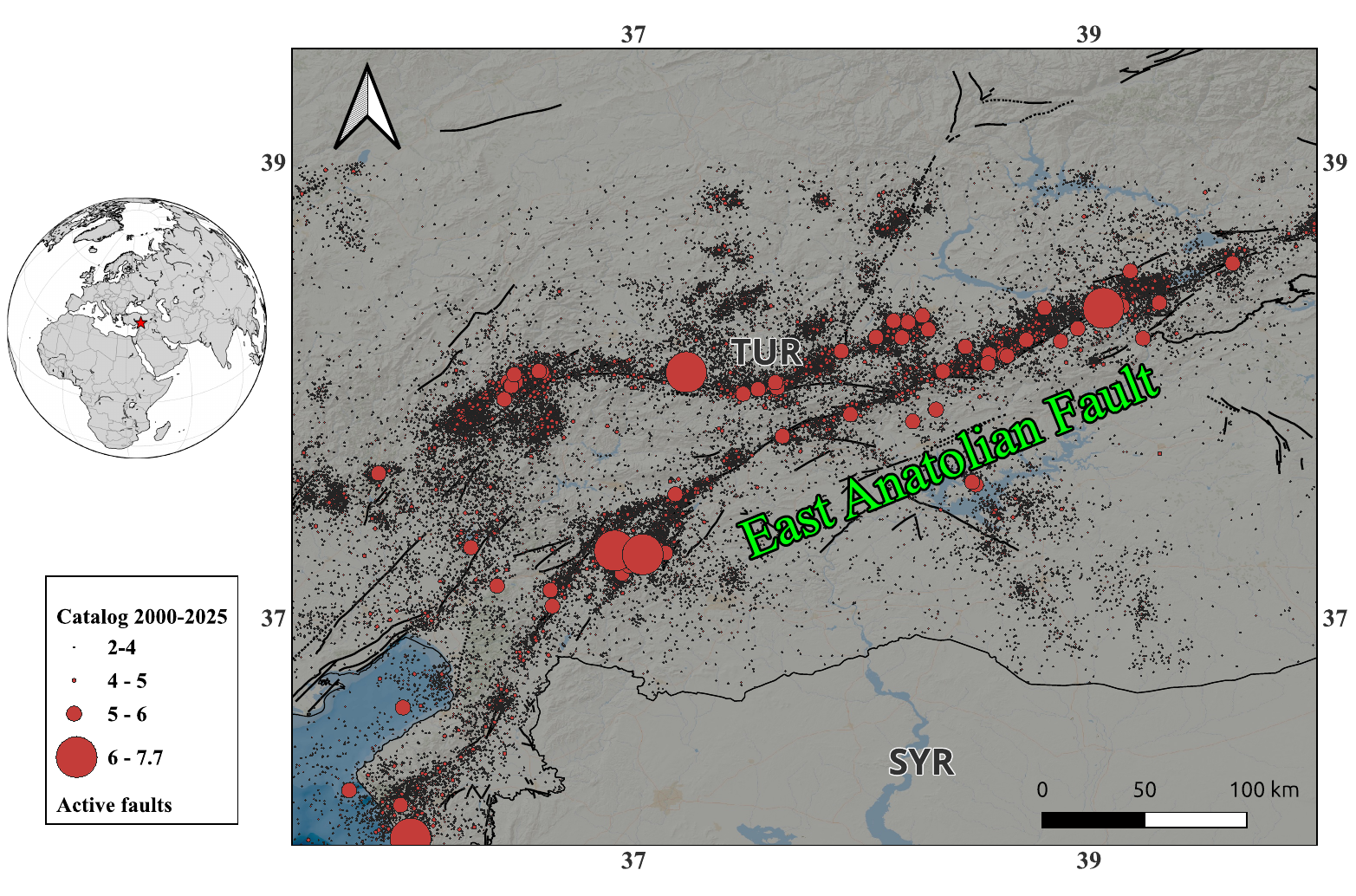}
    \caption{Seismotectonic map of the study area. The main panel shows the spatial distribution of regional seismicity ($M_w \ge 2.0$) from 2000 to 2025 and the active fault systems of the East Anatolian Fault Zone. The inset map provides the regional tectonic context of the Anatolian Plate and surrounding boundaries, outlining the geographic location of the investigated area.}
    \label{fig:fig1}
\end{figure}

This investigation will employ an exhaustive seismic catalog with 45,568 events, which was obtained from the reputable Disaster and Emergency Management Authority of Turkey (AFAD). The seismic catalog covers a time period of 25 years from January 13, 2000, up to October 7, 2025. It is considered complete for $M_w \ge 2.0$, hence providing a strong statistical basis for further analysis. Each event in the seismic catalog includes vital parameters such as origin time, latitude, longitude, focal depth, and magnitude. Before analysis, the raw seismic catalog was subjected to intensive preprocessing. In particular, duplicate events were eliminated, while the declustering process was carried out using the Gardner-Knopoff algorithm \citep{gardner1974} with optimal parameters to separate independent main shocks from their respective foreshock and aftershock clusters. The refined data set therefore offers high spatial and temporal resolution, which in turn provides a reliable basis for the comprehensive characterization of regional seismicity, including the determination of the recurrence relations of earthquakes, the examination of spatial and temporal clustering, and the determination of the structure of active faults. The distribution of seismicity and active faults defining the tectonic framework of the region is illustrated in Fig. \ref{fig:fig1}.

A chronological summary of the declustered and processed seismic activity from 2000 to 2025 is presented in Table \ref{tab:tab1}. The catalog comprises background seismicity alongside major earthquake sequences, highlighting significant temporal fluctuations in event counts, magnitudes, and hypocentral depths. Notably, the year 2023 exhibits an extreme statistical anomaly, recording 20,509 events, which includes the $M_w$ 7.8 mainshock. This surge in seismicity corresponds to the activation of the fault system across a wide depth range (0.21 to 96.41 km), indicating slip propagation from shallow crustal depths down to the upper mantle. Analyzing this major seismic sequence provides critical insights into the co-seismic stress release, aftershock distribution, and overall seismic hazard within the East Anatolian Fault Zone.

\begin{table}[ht!]
\singlespacing
\centering
\footnotesize
\caption{Summary of annual seismic activity, 2000--2025.}
\label{tab:tab1}
\begin{tabular}{cccc}
\hline
\textbf{Year} & \textbf{Number Of Events} & \textbf{Magnitude Range} & \textbf{Depth Range (km)} \\
\hline
2000 & 21 & 2.8--4.8 & 1.0--14.7 \\
2001 & 24 & 2.8--5.1 & 1.2--32.8 \\
2002 & 48 & 2.4--4.8 & 1.0--33.0 \\
2003 & 144 & 2.4--5.7 & 1.0--32.0 \\
2004 & 551 & 2.0--5.3 & 1.0--42.7 \\
2005 & 525 & 2.1--5.2 & 1.0--42.0 \\
2006 & 355 & 2.0--4.2 & 1.0--28.3 \\
2007 & 1270 & 2.0--5.4 & 1.0--75.7 \\
2008 & 1690 & 2.0--4.9 & 1.01--36.02 \\
2009 & 2122 & 2.0--5.0 & 1.03--72.94 \\
2010 & 2822 & 2.0--5.1 & 1.01--33.83 \\
2011 & 558 & 2.0--5.3 & 1.22--29.96 \\
2012 & 692 & 2.0--5.1 & 1.32--40.26 \\
2013 & 2168 & 2.0--4.7 & 1.24--44.89 \\
2014 & 495 & 2.0--4.5 & 0.0--51.37 \\
2015 & 636 & 2.0--5.0 & 0.19--24.3 \\
2016 & 485 & 2.0--4.1 & 1.05--45.42 \\
2017 & 902 & 2.0--5.5 & 0.1--36.57 \\
2018 & 567 & 2.0--5.1 & 0.0--84.46 \\
2019 & 693 & 2.0--5.2 & 2.11--40.36 \\
2020 & 2214 & 2.0--6.8 & 2.25--48.94 \\
2021 & 636 & 2.0--4.7 & 2.71--48.77 \\
2022 & 600 & 2.0--5.2 & 1.97--34.52 \\
2023 & 20509 & 2.0--7.8 & 0.21--96.41 \\
2024 & 3394 & 2.0--5.9 & 0.53--28.5 \\
2025 & 1447 & 2.0--4.9 & 0.46--25.86 \\
\hline
\end{tabular}
\end{table}

\subsection{GRACE and GRACE-FO Data}\label{subsec:grace}

The data for this analysis consist of the month-wise mass concentration solution (RL06.3, version RL0603rc24cE) from the GRACE and GRACE-FO Release 06 Level-3 products, obtained from the University of Texas Center for Space Research. This solution provides the Equivalent Water Height (EWH). This measure represents a vertical integration of the total change in water storage, expressed in centimeters (cm), and is equivalent to the Liquid Water Equivalent Thickness (LWE Thickness) \citep{Tapley_2004, Scanlon_2016, Scanlon_2018}. This dataset is global in scope and facilitates the analysis of changes in terrestrial water storage derived from both the Gravity Recovery and Climate Experiment (GRACE) and its successor mission, GRACE-FO. The data set offers a continuous record spanning from April 2002 to May 2025, incorporating measurements obtained from both missions: GRACE (April 2002 to June 2017) and GRACE-FO (June 2018 to May 2025). This duration encompasses 245 monthly solutions, which includes the anticipated 12-month operational hiatus between the two missions. In terms of spatial representation, the data is provided on an equidistant geographical grid with cell dimensions of $1^\circ \times 1^\circ$, ensuring comprehensive global coverage from $89.875^\circ\text{N}$ to $89.875^\circ\text{S}$ latitude. It is crucial to note that despite the high-density grid spacing, the inherent spatial resolution for the GRACE/GRACE-FO signal is approximately 300 km, as the mascon solutions are effectively resolved on a 1-degree equal area grid. The data underwent essential geophysical corrections and post-processing procedures conducted by the vendor: A spatial mean field from January 2004 to December 2009 was subtracted to represent the data as anomalies. The signal associated with Glacial Isostatic Adjustment (GIA) was removed using the ICE-6G\_D (VM5a) model \citep{Peltier_2018}. Corrections for geocenter motion were applied following the TN-13 formulation. The degree-one spherical harmonic coefficients (geocenter motion) were adjusted using values from \citet{Swenson_2008} for replacement. Additionally, the C20 and C30 coefficients were substituted with more accurate estimates derived from Satellite Laser Ranging (SLR) \citep{Loomis_2020, Ga_dyn_2024}. Outputs from the atmospheric and oceanic de-aliasing model (RL06 AOD1B) were restored for oceanic pixels. An ellipsoidal correction was implemented, aligning the grid with the WGS84 ellipsoid. The final time series of equivalent water heights for the study areas was derived from this processed global dataset for further analysis. It is important for users to be aware of a documented decline in data quality following August 2016 for GRACE and the GRACE-FO-C2 instrument, which is attributed to its dependence on transplanted accelerometer data. This research investigates a 4-month timeframe, encompassing both the period preceding and following the Kahramanmara\c{s} earthquake in February 2023. Consequently, only GRACE-FO data was utilized. Geodetic processing and gravity-field analysis are crucial for interpreting satellite gravimetry data, often supported by open-source Python toolboxes designed for datum unification and vertical analysis \citep{Kelly2026}. Furthermore, modern research increasingly relies on web-oriented geospatial software systems for integrated access and preliminary analysis of geological and geophysical data \citep{Soloviev2026}. Spatiotemporal analysis of regional water bodies and terrestrial water storage changes has also been successfully demonstrated in adjacent Middle Eastern basins to track long-term environmental and hydrological variations \citep{Li2026}.

\section{Result and Discussion}\label{sec:results}

\begin{figure}[ht!]
    \centering
    \includegraphics[width=\textwidth]{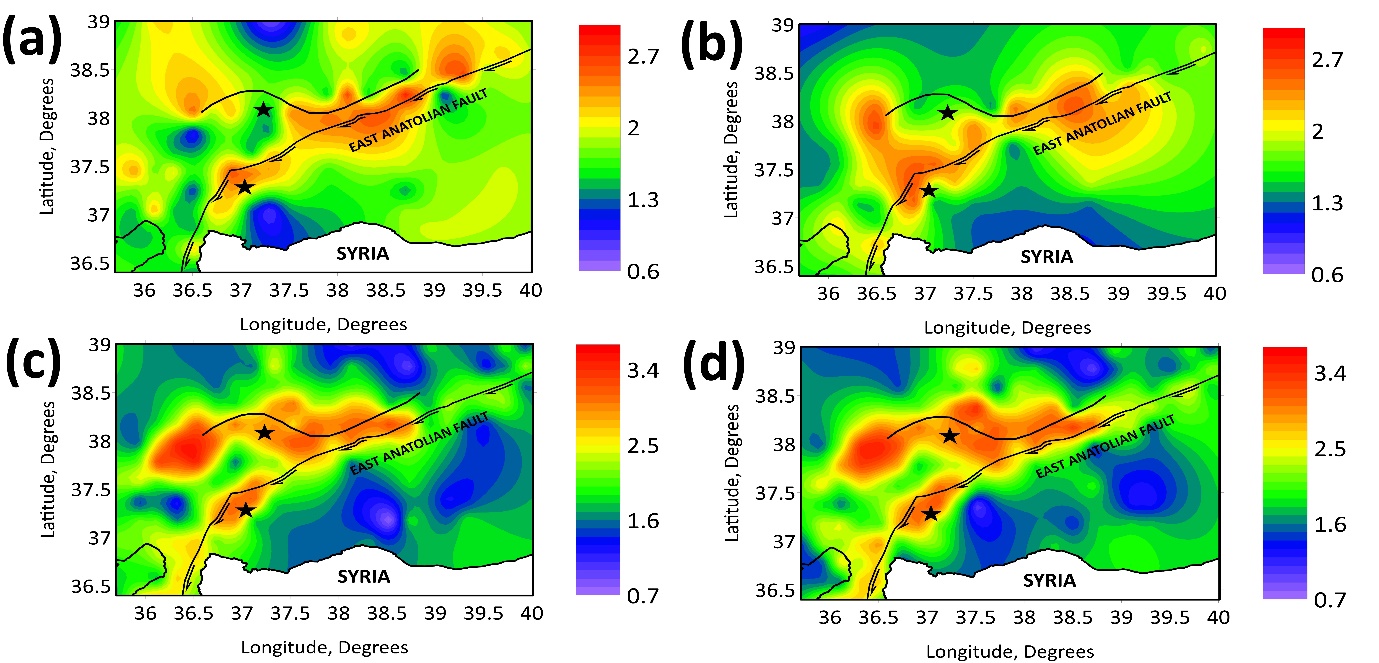}
    \caption{Spatial variation of Shannon entropy (a) 2.5 years before, (b) 1 year before, (c) 1 year after, and (d) 2.5 years after relative to the 6 February 2023 Kahramanmara\c{s} earthquakes.}
    \label{fig:fig2}
\end{figure}

The spatial variation of Shannon entropy across the East Anatolian Fault system shows an unmistakable transformation of the earthquake patterns as the crust prepared for the 6 February 2023 Kahramanmara\c{s} earthquakes (Fig. \ref{fig:fig2}). During the 2.5-year pre-seismic period, the values of Shannon entropy demonstrate uniformity, indicating that the seismic activity was stable with earthquake patterns following a predictable spatial distribution. This indicates that the fault network was in a coherent state, meaning that the earthquakes had occurred with relatively homogeneous stress distribution. However, during the last year of the pre-seismic period, the spatial progression of Shannon entropy shows an unmistakable rise in the values of Shannon entropy around the future earthquake locations. This rise indicates the breakdown of spatial uniformity and the rise of localized complexities, indicating that the fault network was transitioning into a disordered state. Such patterns can be associated with the buildup of stress, the activation of small fractures, and the development of micro-clusters that destabilize the earthquake patterns. The choice of spatial and temporal parameters, such as window smoothing and grid sizing, is known to significantly affect the stability and resolution of calculated seismic parameters during large earthquake preparatory and coseismic cycles \citep{ZhangX2025}. After the main earthquake sequence, the values of Shannon entropy have become heterogeneous with high values of Shannon entropy along the ruptured faults and structures. These patterns can be associated with the strong activity of aftershocks and the reorganization of the fault network. Also in 2.5 years after the earthquake, the values of Shannon entropy remain high around the ruptured faults, indicating that the earthquake has had a lasting effect on the stress distribution of the region.

The Tsallis entropy offers a more sophisticated and physically meaningful perspective on the non-extensiveness and long-range interactions inherent in the seismic system (Fig. 3). During the 2.5-year pre-seismic period, the value of the Tsallis entropy remains moderate, indicating that the system was primarily governed by local interactions and weak long-range coupling between fault segments. This behavior suggests that, although the crust was accumulating stress, the interactions between seismic elements had not yet evolved into a strongly correlated state. As the earthquake approaches, the Tsallis entropy exhibits a marked and spatially coherent increase in the vicinity of the earthquake epicenter. This increase is more pronounced and more sharply defined than the corresponding rise in the Shannon entropy, demonstrating the superior capability of the Tsallis framework to capture long-range correlations and cooperative effects among seismic system components. Such behavior is typical of systems approaching criticality, where interactions extend beyond local neighborhoods and the fault system begins to act as a coupled multi-scale structure. The sensitivity of Tsallis entropy to non-extensive effects enables it to detect the growing connectivity of the fault system, indicating that the crust was approaching a state in which small perturbations could propagate over long distances.

In the post-mainshock period, the Tsallis entropy reveals a distinctly fragmented and elongated pattern aligned with the ruptured fault segments. This pattern effectively captures both the activation of secondary fault structures and the redistribution of stress throughout the region. The elevated values of Tsallis entropy during the 2.5-year post-seismic interval indicate that the system remains in a non-equilibrium state, with long-range interactions continuing to dominate. Such persistent non-extensive behavior is consistent with theoretical expectations for large continental strike-slip earthquakes, where fault-zone damage and stress transfer can sustain long-range interactions long after the mainshock. To quantitatively characterize the statistical properties of the seismicity, the $b$-value, magnitude of completeness ($M_c$), and the Tsallis parameter $q$ were determined using maximum likelihood estimation and uncertainty propagation, as detailed in Algorithm 2. The MAXC analysis yielded a $b$-value of $0.83 \pm 0.04$ and a $M_c$ of 2.5. This $b$-value, which is markedly lower than the global average of approximately 1.0 \citep{frohlich1993}, indicates a stress regime dominated by locked asperities and elevated differential stress. Such conditions are consistent with the tectonic characteristics of the Kahramanmara\c{s} region, where complex fault interactions and continuous stress accumulation promote the occurrence of major seismic events. Substitution of the obtained $b$-value into the analytical relation results in a Tsallis entropy index of $q = 1.70 \pm 0.02$, fully consistent with the non-extensive cumulative magnitude distribution.

\begin{algorithm}[htbp]
\singlespacing
\caption{Relationship Between Tsallis q and Gutenberg--Richter b-value}
\label{alg:tsallis_gr_relation}
\begin{algorithmic}[1]\footnotesize
\Require{magnitude vector $M = \{M_1, M_2, \dots, M_n\}$, bin width $\Delta m$}
\Statex \textbf{Phase 1 --- Estimate $b$-value (MLE) \& Uncertainty}
\State $\text{bins} \gets \text{ARANGE}(\min(M), \max(M) + \Delta m, \Delta m)$
\State $\text{counts} \gets \text{HISTOGRAM}(M, \text{bins})$, $M_c \gets \text{centre of bin with } \text{ARGMAX}(\text{counts})$
\State $M_c \gets M[M \ge M_c]$, $M_{\text{min}} \gets M_c - \Delta m/2$
\State $b \gets \log_{10}(e) / (\text{MEAN}(M_c) - M_{\text{min}})$
\State $\sigma_b \gets 2.30 \cdot b^2 \cdot \text{STD}(M_c) / \sqrt{|M_c|}$
\Statex \textbf{Phase 2 --- Derive Tsallis $q$ \& Uncertainty Propagation}
\State $q \gets (b + 4) / (b + 2)$
\State $\sigma_q \gets | -2 / (b + 2)^2 | \cdot \sigma_b$
\Statex \textbf{Phase 3 --- Physical Interpretation \& Consistency Check}
\If{$q < 1$}
    \State \textbf{warn} ``sub-extensive regime (rare in seismicity)''
\EndIf
\If{$q = 1$}
    \State \textbf{note} ``reduces to Boltzmann--Gibbs ($b \to \infty$)''
\EndIf
\If{$1 < q < 2$}
    \State \textbf{note} ``super-extensive; typical seismic range''
\EndIf
\If{$q \ge 2$}
    \State \textbf{warn} ``non-normalizable distribution; check $b$''
\EndIf
\Statex \textbf{Phase 4 --- Non-Extensive Cumulative Distribution Fit}
\State $P(\ge M) \gets \left[ 1 - \frac{(1-q)(M - M_c)}{2-q} \right]^{\frac{1}{1-q}}$
\State $N_{\text{obs}}(\ge M) \gets \text{COUNT}(M_c \ge M)$ for $M \in \text{ARANGE}(M_c, \max(M_c), \Delta m)$
\State $N_{\text{fit}}(\ge M) \gets N_{\text{total}} \cdot P(\ge M)$
\State $\text{RMSE} \gets \sqrt{ \text{MEAN}( (\log_{10}(N_{\text{obs}}) - \log_{10}(N_{\text{fit}}))^2 ) }$
\Statex \textbf{Output}
\State \Return $\{ M_c, b, \sigma_b, q, \sigma_q, \text{RMSE} \}$
\State Print ``$b = b \pm \sigma_b \to q = q \pm \sigma_q$''
\end{algorithmic}
\end{algorithm} The magnitude of q, significantly exceeding unity, underscores the necessity of employing Tsallis entropy instead of classical Boltzmann--Gibbs statistics for characterizing the investigated system. The highly non-extensive and non-equilibrium nature of the seismic process is thus firmly established by the value of q, reinforcing the interpretation that the Kahramanmara\c{s} seismicity is governed by long-range interactions, cooperative dynamics, and multi-scale fault-system behavior.

\begin{figure}[ht!]
    \centering
    \includegraphics[width=\textwidth]{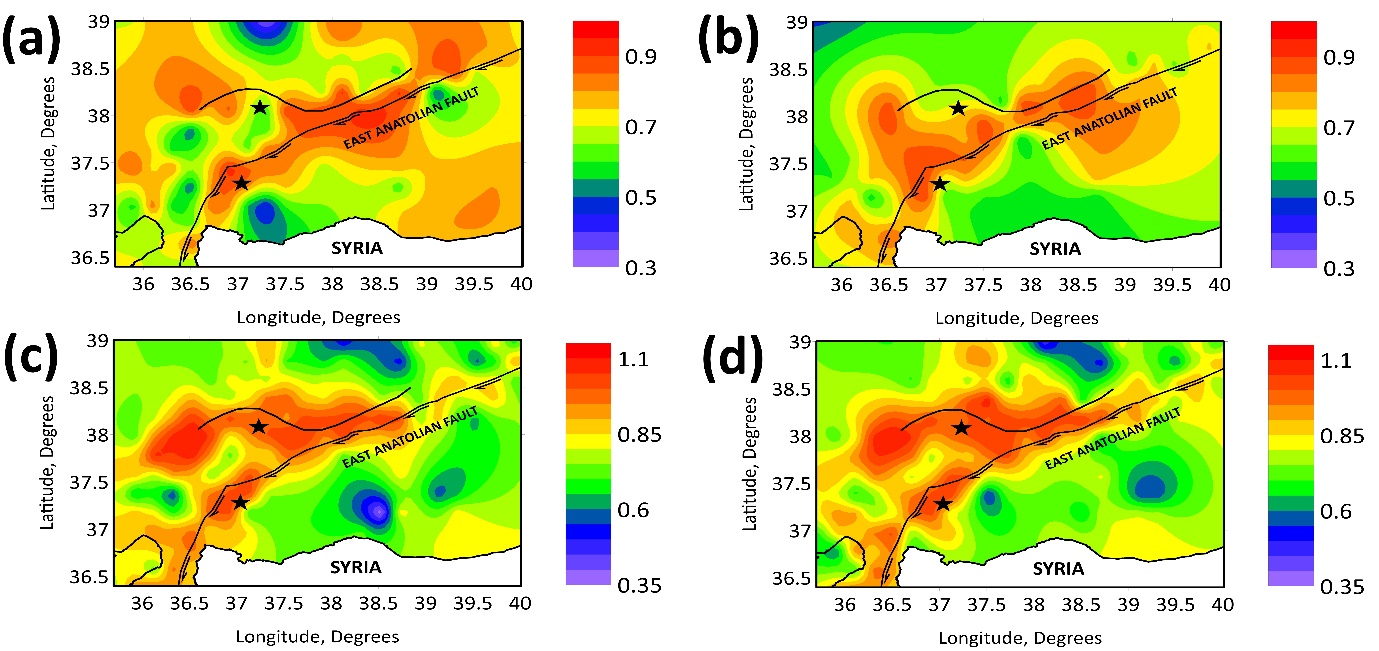}
    \caption{Spatial variation of Tsallis entropy (a) 2.5 years before, (b) 1 year before, (c) 1 year after, and (d) 2.5 years after relative to the 6 February 2023 Kahramanmara\c{s} earthquakes.}
    \label{fig:fig3}
\end{figure}

The Equivalent Water Height (EWH) fields provided by the GRACE-FO data indicate a clear pattern of short-term mass redistributions during the months preceding the Kahramanmara\c{s} earthquakes, offering an alternative geodetic view on the dynamic state of the crust (Fig. 4). In the months preceding the earthquake, four and two months before the earthquake, moderate negative anomalies are seen near the central East Anatolian Fault, indicating gradual mass loss related to hydrological depletion, poroelastic contraction, and minor crustal deformation. These early anomalies indicate that the crust was already responding to the growing stress with noticeable adjustments. One month before the earthquake, a more pronounced negative anomaly is seen near the future earthquake location, indicating that the crust was preparing for an episode of extreme sensitivity. This anomaly could be related to the contraction of pore space, the migration of fluids along newly created micro-fractures, and the onset of elastic deformation as a result of stress accumulation. In the immediate aftermath of the earthquake, the EWH field undergoes a sudden shift, marked by the development of negative anomalies consistent with coseismic elastic deformation and rapid mass redistribution. The spatial pattern of these anomalies indicates that the earthquake triggered a large-scale reorganization of the crust, influencing both the mechanical and hydrological properties of the region. In the two and four months following the earthquake, the anomalies gradually diminish but remain detectable near the earthquake locations. These could be related to the post-earthquake relaxation and related responses in the crust and upper mantle, including the effects of fault zone damage and the resulting permeability enhancements.

\begin{figure}[ht!]
    \centering
    \includegraphics[width=\textwidth]{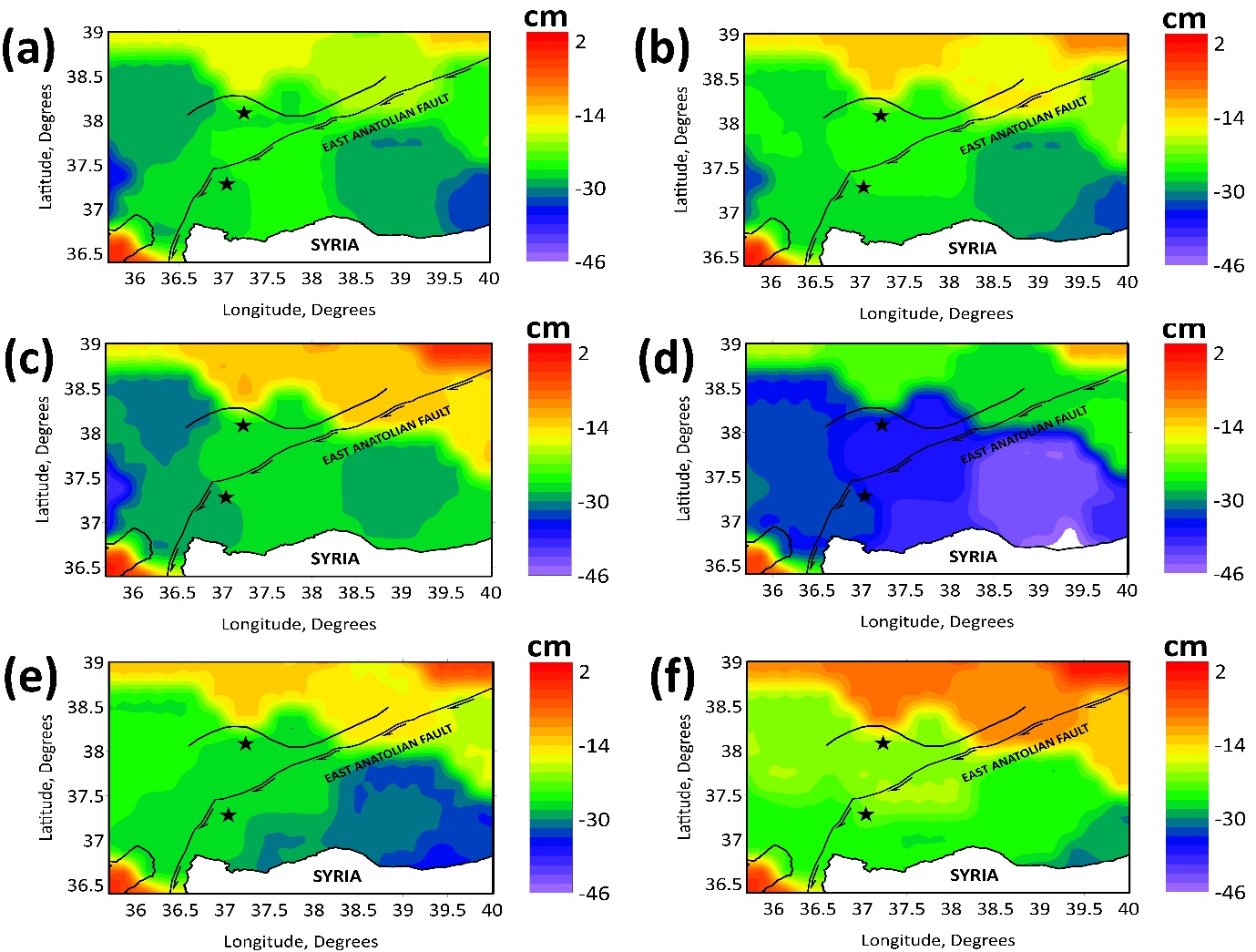}
    \caption{Spatial variation of GRACE-FO terrestrial water storage anomalies (a) 4 months before, (b) 2 months before, (c) 1 month before, (d) 1 month after, (e) 2 months after, and (f) 4 months after the 2023 Kahramanmara\c{s} earthquakes.}
    \label{fig:fig4}
\end{figure}

The joint variation of Shannon entropy, Tsallis entropy, and EWH indicates the close interrelation of the processes that controlled the preparation, rupture, and recovery of the East Anatolian Fault System. Before the earthquake occurred, the two entropy values displayed an increasing trend, which indicates the increasing level of disorganization and structural complexity of the seismic field. This increasing trend also indicates the fragmentation of the fault network, the development of micro-cracks, and the strengthening of stress interactions at different scales. Concurrently, the GRACE-FO data revealed the continuous decrease of the EWH values, which is associated with mass redistribution due to poro-elastic contraction, fluid migration, and minor crustal deformation. However, during the mainshock sequence, this relationship is even clearer. The sudden jump in both Shannon and Tsallis entropy levels serves as an indicator of the explosive development of seismic complexity during the aftershock cascade. At the same time, the sudden drop in EWH levels corresponds to the rapid redistribution of mass during the earthquake. The simultaneous response of all three parameters serves as an indicator of the extent to which this earthquake has restructured the system, changing both its mechanical and hydrological states. During the post-seismic period, the maintenance of high levels of entropy near the earthquake regions mirrors the maintenance of anomalies in EWH levels. This serves as an indicator of the extent to which mechanical and hydrological recovery processes are related. Processes such as fault zone healing, stress redistribution, fluid migration, and viscoelastic relaxation all play an important role in determining the system's evolution. The similarity between patterns of entropy and EWH serves as an indicator of the extent to which seismic complexity, long-range fault interactions, and mass redistribution are related. All three of these parameters serve as an indicator of the multi-dimensional nature of earthquake processes. They demonstrate that all three phases of earthquake processes pre-seismic, co-seismic, and post-seismic are related.

\section{Conclusion}\label{sec:conclusion}

This research demonstrates that the history of the East Anatolian Fault System in the period before the Kahramanmara\c{s} event on 6 February 2023 is not simply contained in the location and frequency of the earthquakes. Rather, it can also be seen in its statistical complexity, its interplay of faults over long distances, and its hidden rearrangement of mass in the short term. Through the combination of a 25-year high-resolution seismic catalog, GRACE-FO mass change data, and two entropy measures, we observe the system's steady progression toward a tipping point in the months prior to the event. This is seen in the coordinated rise of Shannon entropy and Tsallis entropy, in which order in space gives way to multiscaling and clustering, as nonlocal interactions increase, indicating a progress toward instability in the fault system. Concurrently, the regular decrease in EWHT points to poroelastic contraction, fluid flow, and deformation, which might mirror the growing disorder in the system. Within the main shock sequence, the steep rise of both entropy metrics with the corresponding steep decline in EWH indicates a prompt rearrangement of the crust as stress and mass redistributions occurred rapidly over the region. Also, the elevated levels of entropy remaining for 2.5 years following the main event indicate that the system failed to revert to equilibrium conditions; instead, it remained in a non-equilibrium position for an extended period as a result of aftershocks, damaged fault zones, higher permeability, and relaxation effects of viscoelasticity. The joint variation of these three independent indicators statistical entropy, non-extensive interactions, and satellite-derived mass change provides a multidimensional view of the pre-seismic, co-seismic, and post-seismic phases of a major continental strike-slip earthquake. The research demonstrates the potential of fusing information theory with spaceborne measurement of mass change to detect critical transition points on active fault systems. The fusion may hold promise for improved monitoring and risk assessment of tectonically active regions globally.

\section*{Acknowledgment}

We gratefully acknowledge the Turkey's Disaster and Emergency Management Authority (AFAD) for providing the comprehensive earthquake catalog used in this study, which was instrumental in conducting the spatiotemporal analysis of seismicity.

\section*{Author Contributions}

Muhammed Hossein Mousavi: Software, Data curation, Visualization, Writing - original draft

Hamzeh Mohammadigheymasi: Conceptualization, Supervision, Methodology, Writing - review \& editing

Amirreza Moradi: Validation, Formal analysis

\section*{Funding Information}

The authors declare that no funds, grants, or other support were received during the preparation of this manuscript.

\section*{Declaration of Competing Interest}

The authors declare that they have no known competing financial interests or personal relationships that could have appeared to influence the work reported in this paper.

\section*{Data Availability}

The earthquake catalogue is provided by the Disaster and Emergency Management Authority of Turkey (AFAD) and can be accessed at \url{https://udim.afad.gov.tr/}. The monthly Equivalent Water Height (EWH) datasets from the GRACE and GRACE-FO missions are publicly available from the University of Texas Center for Space Research (CSR) at \url{https://www2.csr.utexas.edu/grace/}.

\bibliographystyle{elsarticle-harv}
\bibliography{mybibfile}

\end{document}